\documentstyle[sprocl]{article}

\input{psfig}

\def\be{\begin{equation}}
\def\ee{\end{equation}}
\def\bea{\begin{eqnarray}}
\begin{document}

\title{  THE LOGARITHMIC DERIVATIVE OF THE $F_{2}$ STRUCTURE
FUNCTION AND SATURATION}

\author{E. GOTSMAN}

\address{School of Physics and Astronomy, Tel Aviv University,\\
Tel Aviv 69978,  Israel\\E-mail: gotsman@post.tau.ac.il}

%\frenchspacing
%\renewcommand{\topfraction}{.99}
%\renewcommand{\textfraction}{.01}
%\addtolength{\textheight}{.2cm}

\maketitle\abstracts{ We show that when screening corrections are included
$\frac{\partial F_{2}(x,Q^{2})}{\partial ln(Q^{2}/Q_{0}^{2})}$
is consistent with the behaviour that one expects in pQCD. Screening
corrections explain the enigma of the Caldwell plot. Saturation has not
 been reached at present HERA energies.}

\section{Introduction}
   The Caldwell plot \cite{Cald} of 
 $\frac{\partial F_{2}(x,Q^{2})}{\partial ln(Q^{2}/Q_{0}^{2})}$
presented at the Desy Workshop in November 1997 suprized the community.
The results appeared to indicate that we have reached a region in the 
x and $Q^{2}$ where pQCD was no longer valid.
DGLAP evolution leads us to expect that
 $\frac{\partial F_{2}(x,Q^{2})}{\partial ln(Q^{2}/Q_{0}^{2})}$
at fixed $Q^{2}$ would be a monotonic increasing function of 
$\frac{1}{x}$, whereas a superficial glance at the data suggests that the
logarithmic derivative of $F_{2}$ deviates from the expected pQCD
behaviour, and has a turnover in the region of
2 $ \leq  Q^{2} \leq $ 4 GeV$^{2}$ (see fig.1 where the ZEUS data
and the GRV'94
predictions are shown).
 Opinions were  voiced that the phenomena was
connected with the transition from "hard" to "soft" interactions. 
Others \cite{AHM} felt that the "turnover" in
$\frac{\partial F_{2}(x,Q^{2})}{\partial ln(Q^{2}/Q_{0}^{2})}$
may be an indication of saturation of the parton distributions.

\begin{figure}[p]
%\rule{5cm}{0.2mm}\hfill\rule{5cm}{0.2mm}
%\vskip 2.5cm
%\rule{5cm}{0.2mm}\hfill\rule{5cm}{0.2mm}
\hspace*{0.4cm}\psfig{figure=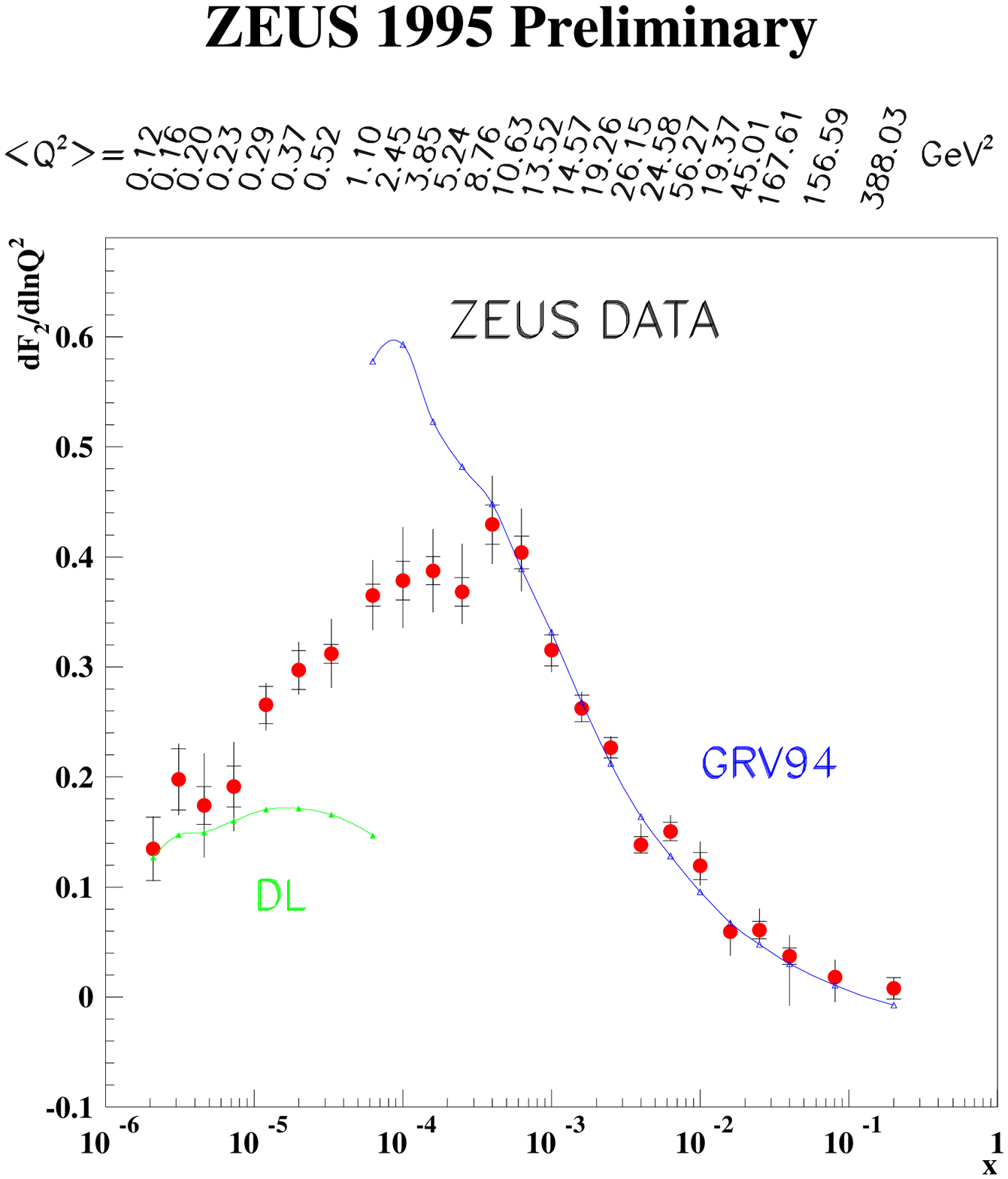,width=10cm,height=2.5in}
%\vspace*{2mm}
\caption{ ZEUS data and GRV'94 predictions for $F_{2}$ slope
\label{fig:fig.1}}
 \end{figure}

\par Amongst the problems that one faces in attempting to comprehend the
data, is the fact that due to kinematic constraints the data is sparse,
and each point shown pertains to a different pair of values of x and
$Q^{2}$. We miss the luxury of having measurements at several different
values of x for  fixed values of $Q^{2}$, which would allow one to deduce
the
detailed behaviour of
 $\frac{\partial F_{2}(x,Q^{2})}{\partial ln(Q^{2}/Q_{0}^{2})}$.
Bartels et al \cite{BCF} had  previously suggested that the logarithmic
derivative of the structure function $F_{2}$ should be sensitive to
screening effects.

\section{QCD and 
$\frac{\partial F_{2}(x,Q^{2})}{\partial ln(Q^{2}/Q_{0}^{2})}$}

 The DGLAP evolution equations \cite{DGLAP} imply the relation
\be
\frac{\partial F_{2}(x,Q^{2})}{\partial ln(Q^{2}/Q_{0}^{2})}\;= \;
\frac{2 \alpha s}{9 \pi} xG^{DGLAP}(x,Q^{2})
\ee
where $xG^{DGLAP}(x,Q^{2})$ denotes the distribution of gluons in the
proton. At present HERA energies $xG(x,Q^{2})$   grows rapidly
 with increasing  $\frac{1}{x}$ i.e. energy. From unitarity constraints
we know that this growth must taper off, and at some value of 
$x$ (=$x_{cr}$), $xG(x,Q^{2})$ must become saturated, and perturbative QCD
will no longer be valid.

\par To illustrate the effects that we can expect for the saturated case,
we
turn to the colour dipole picture of DIS \cite{AHM}. Here the $\gamma^{*}$
fluctuates
into a $q\bar{q}$ pair, which then scatters on the proton over a
relatively short time scale compared to the fluctuations.
We have
\be
\frac{\partial F_{2}(x,Q^{2})}{\partial ln(Q^{2}/Q_{0}^{2})}
\; \sim \; Q^{2} \sigma_{q\bar{q}}(\Delta r_{\bot} \sim \frac{1}{Q})
\ee
where $\sigma_{q\bar{q}}$ denotes the cross section for the $q\bar{q}$
pair to interact with the proton, and $\Delta r_{\bot}$
the distance between the q and $\bar{q}$.
When the distribution of gluons in the proton is normal pQCD is
applicable, and the relation given in eq.(1) holds. However, when the
gluons are densely packed (i.e. saturated) one reaches the unitarity limit
and the colour dipole cross section can be assumed to be geometric
i.e. $\sigma_{q\bar{q}} \sim \pi R_{p}^{2} $.

\par For the saturated case we then expect
\be
\frac{\partial F_{2}(x,Q^{2})}{\partial ln(Q^{2}/Q_{0}^{2})} \; \sim
\; Q^{2} \pi R_{p}^{2}
\ee
i.e. the logarithmic slope should grow linearly with $Q^{2}$.

\section{Results}
\par We show that the Caldwell plot is in agreement with the pQCD
expectations, once screening corrections (SC) (which become more important
as one goes to lower values of x and $Q^{2}$), are included. To provide a
check of our calculations, we compare with the results one derives
using the  ALLM'97  parametrization \cite{ALLM}, which we use as a "pseudo
data base". This parameterization is based on a Regge-type approach
formulated so as to be compatible with pQCD and the DGLAP evolution
equations.

\begin{figure}[p]
%\rule{5cm}{0.2mm}\hfill\rule{5cm}{0.2mm}
%\vskip 2.5cm
%\rule{5cm}{0.2mm}\hfill\rule{5cm}{0.2mm}
\hspace*{0.5cm}\psfig{figure=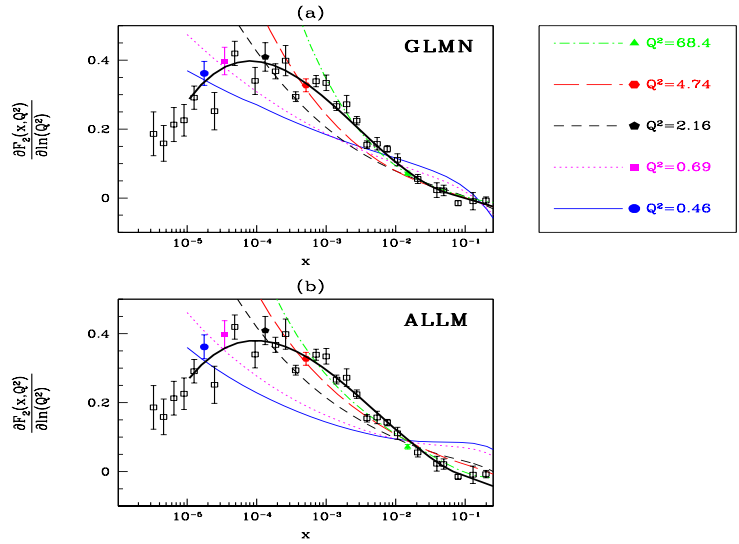,width=10cm,height=3.in}
\caption{ The  $F_{2}$ slope (a)  in our QCD calculation 
incorporating SC, and (b) in the ALLM"97 parametrization. \label
{fig:fig.2}}
 \end{figure}

\par Following the method suggested by Levin and Ryskin \cite{LR} and
Mueller \cite{M1} we calculate the SC pertaining to 
 $\frac{\partial F_{2}(x,Q^{2})}{\partial ln(Q^{2}/Q_{0}^{2})}$
for both the quark and gluon sector.  In fig.2 we show the results
as well as those of ALLM compared with the experimental results.

\begin{figure}[t]
%\rule{5cm}{0.2mm}\hfill\rule{5cm}{0.2mm}
%\vskip 2.5cm
%\rule{5cm}{0.2mm}\hfill\rule{5cm}{0.2mm}
\psfig{figure=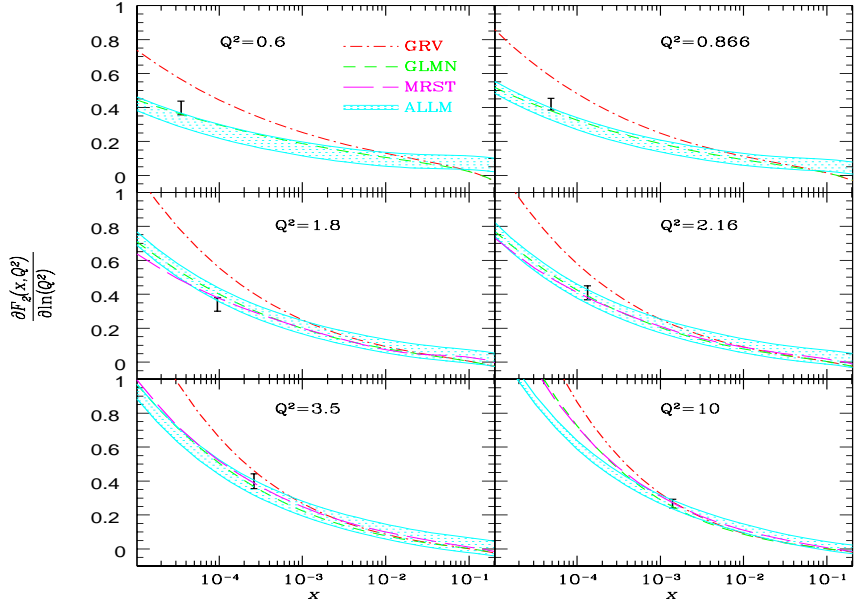,width=12cm,height=3.5in}
%\vspace*{2mm}
\caption{ $\frac{\partial F_{2}(x,Q^{2})}{\partial ln(Q^{2}/Q_{0}^{2})}$.
 In addition to the ALLM band we show a typical data point with its
error. \label{fig:fig3.}}
 \end{figure}

\begin{figure}[t]
%\rule{5cm}{0.2mm}\hfill\rule{5cm}{0.2mm}
%\vskip 2.5cm
%\rule{5cm}{0.2mm}\hfill\rule{5cm}{0.2mm}
\psfig{figure=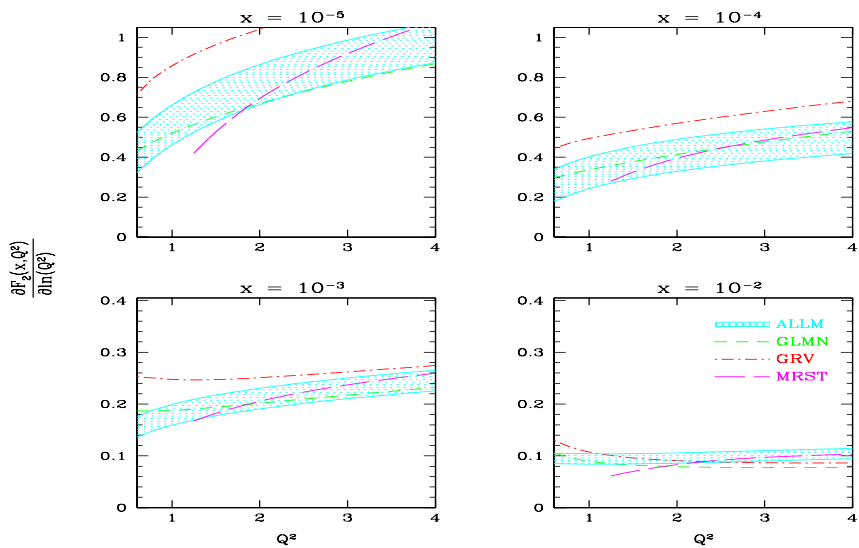,width=12cm,height=3.in}
%%\vspace*{2mm}
\caption{
 $\frac{\partial F_{2}(x,Q^{2})}{\partial ln(Q^{2}/Q_{0}^{2})}$
at fixed x. \label{fig:fig.4}}
\end{figure}

\par In fig.3 and 4  
we display our calculations for the logarithmic derivative
of $F_{2}$ after SC have been incorporated, as well as the ALLM results.
In fig.3 for fixed values of $Q^{2}$ and varying values of x, and in
fig.4 for fixed x and varying values of $Q^{2}$. In fig.4 we show our
results as well as those of ALLM compared with the experimental results.
We note that
 $\frac{\partial F_{2}(x,Q^{2})}{\partial ln(Q^{2}/Q_{0}^{2})}$
at fixed $Q^{2}$ both in our calculations and in the "psuedo data" (ALLM),
remains a $ \bf{monotonic}$ increasing function of $\frac{1}{x}$.  

From fig.4 we note that for fixed x, 
 $\frac{\partial F_{2}(x,Q^{2})}{\partial ln(Q^{2}/Q_{0}^{2})}$
decreases as $Q^{2}$ becomes smaller. The decrease becomes stronger as we
go to lower values of x. This phenomena which is due to SC adds to the
confusion in interpreting the Cadwell plot.

\section{Conclusions}

$\;\;\;\;\;$ 1) We have obtained a good description of 
 $\frac{\partial F_{2}(x,Q^{2})}{\partial ln(Q^{2}/Q_{0}^{2})}$
for x $ \leq $  0.1.

2) Our results suggest that there is a smooth transition between the
"soft" and "hard" processes.

3) SC are essential for describing the Caldwell plot even at present HERA
energies where we are far below the saturation region ( as
$\frac{\partial F_{2}(x,Q^{2})}{\partial ln(Q^{2}/Q_{0}^{2})}$
 $ < F_{2}(x,Q^{2}$)). In the saturation region ($x \leq x_{cr}$) we
expect
$\frac{\partial F_{2}(x,Q^{2})}{\partial ln(Q^{2}/Q_{0}^{2})}$
= $F_{2}(x,Q^{2}$). 

4) At fixed $x$ and/or fixed $Q^{2}$, SC to do not change the qualitative
behaviour of
 $\frac{\partial F_{2}(x,Q^{2})}{\partial ln(Q^{2}/Q_{0}^{2})}$,
they only produce a smaller value of the slope. 

5) The apparent turn over of 
 $\frac{\partial F_{2}(x,Q^{2})}{\partial ln(Q^{2}/Q_{0}^{2})}$
is an illusion, created by the experimental limitation in measuring the
logarithmic derivative of $F_{2}$ at particular correlated values of
$Q^{2}$ and x.

6) Direct experimental evidence supporting  our hypothesis, was
presented recently
by Max Klein at the LP99 conference \cite{MK}. He concluded "H1 see no
departure from the rising behaviour of
$\frac{\partial F_{2}(x,Q^{2})}{\partial ln(Q^{2}/Q_{0}^{2})}$
as a function of increasing $\frac{1}{x}$ for 
$Q^{2} \geq 3 \;\; GeV^{2}$ ", (this is the lowest value of $Q^{2}$
for which H1 presented data).

The detailed calculations and results that this talk was based on, appear
in \cite{GLM1} and \cite{GLM2}.

\section{Acknowledgements}
 I would like to thank my friends and collegues
Genya Levin and Uri Maor for an enjoyable and fruitful collaboration.
I am also indebted to  John Dainton for drawing my attention to the
new H1 data.
This research was supported in part by the Israeli Science Foundation
founded by the Israel Academy of Science and Humanities.


\section*{References}
\begin{thebibliography}{99}

\bibitem{Cald}
A. Caldwell, Invited talk, DESY Theory Workshop. DESY, October 1997. 

\bibitem{AHM}
A.H. Mueller, hep-ph/9904404 and hep-ph/9902302. 

\bibitem{BCF}
J. Bartels, K. Charchula and J. Feltesse, Proceedings  "Physics at Hera"
, Ed. W. Buchmueller and G. Ingelman, Hamburg 1991, Vol 1, 193.
\bibitem{DGLAP}
V.N. Gribov and L.N. Lipatov, Sov. J. Nucl. Phys 15, 438 (1972);
L.N. Lipatov, Yad. Fiz. 20, 181 (1974);
G. Altarelli and G. Parisi, Nucl. Phys. B 126, 298 (1977);
Yu.L. Dokshitzer, Sov. Phys. JETP 46, 641 (1977).

\bibitem{ALLM}
H. Abramowicz and A. Levy,  DESY 97-251, hep-ph/9712415. 

\bibitem{LR}
 E.M. Levin and M.G. Ryskin, Sov. J. Nucl. Phys. 45, 150 (1987).

\bibitem{M1} A.H. Mueller, Nucl. Phys. B 335, 115 (1990).


\bibitem{MK}
M. Klein, "Structure functions in deep inelastic Lepton-Nucleon
scattering",
plenary talk, Lepton-Photon'99, XIX International Symposium on Lepton and
Photon Interactions at High Energies, Stanford University, August 1999.

\bibitem{GLM1}
E. Gotsman, E. Levin and U. Maor, Phys. Lett. B 425, 369 (1998).

\bibitem{GLM2}
E. Gotsman, E. Levin and U. Maor, Nucl. Phys. B 539, 535 (1999).


 \end{thebibliography}
\end{document}